\documentclass[times]{elsarticle}
\usepackage{moreverb} 
\usepackage[colorlinks,bookmarksopen,bookmarksnumbered,citecolor=red,urlcolor=red,pdfauthor=author]{hyperref}
\usepackage{amsfonts,amsmath,amssymb}
\usepackage{psfrag,pstricks,pst-node}

\usepackage[section]{placeins}
\usepackage{subfigure} 
\usepackage{mathrsfs}
\usepackage{graphicx}
\usepackage[ruled,vlined]{algorithm2e}
\usepackage{fancyhdr}  
\usepackage{psfrag} 
\usepackage[export]{adjustbox} 
\usepackage{enumerate} 
\usepackage{comment}
\usepackage{lmodern}
\usepackage{media9}
\usepackage{hyperref}
\usepackage{algorithmic}
\usepackage{bm}
\usepackage{booktabs}
\usepackage{threeparttable}
\usepackage{listings}
\DeclareUnicodeCharacter{FF0C}{,}
\usepackage[margin=2.5cm]{geometry}
\graphicspath{{./figs/}{./model structure/}}
\DeclareGraphicsExtensions{.pdf,.eps}

\usepackage{pgfplots}
\pgfplotsset{compat=1.13}

\begin{document}
\makeatletter
\def\ps@pprintTitle{}
\makeatother
 
\begin{frontmatter}
\renewcommand{\thefootnote}{\fnsymbol{footnotemark}}

\fancypagestyle{plain}{
\fancyhf{} 
\fancyhead[RO,RE]{\thepage} 
}

\title{A Novel Voting System for Medical Catalogues in National Health Insurance}
    \author[lab2]{Xingyuan Liang\corref{cor1}}      \cortext[cor1]{Corresponding author}     \ead{liangxingyuan5529@link.tyut.edu.cn}
    \author[lab1]{Haibao Wen}
    \address[lab2]{School of Software Engineering, Taiyuan University of Technology, Shanxi 030600，China}
    \address[lab1]{Department of Economics, University of Bath, Bath BA2 7AY, United Kingdom}

\begin{abstract}

This study explores the conceptual development of a medical insurance catalogue voting system. The methodology is centred on creating a model where doctors would vote on treatment inclusions, aiming to demonstrate transparency and integrity. The results from Monte Carlo simulations suggest a robust consensus on the selection of medicines and treatments. Further theoretical investigations propose incorporating a patient outcome-based incentive mechanism. This conceptual approach could enhance decision-making in healthcare by aligning stakeholder interests with patient outcomes, aiming for an optimised, equitable insurance catalogue with potential blockchain-based smart-contracts to ensure transparency and integrity.

\end{abstract}

\begin{keyword}
insurance catalogue, voting system, Monte Carlo Simulations, doctor agent, integrity-driven rating
\end{keyword}
\end{frontmatter}

\section{Introduction}\label{Introduction}
\vspace{-2pt}

In modern healthcare insurance systems, decision-making processes face significant challenges such as opacity and uneven distribution of resources. These issues prevent the timely update of medical insurance catalogues to accurately reflect the latest medical advancements and patient needs \cite{Walshe2009}. Blockchain technology, known for its immutability and distributed ledger capabilities, offers innovative solutions to these problems by ensuring data integrity and transparency, thereby enhancing the fairness and efficiency of decision-making processes in healthcare \cite{Kuo2017}.

Worldwide, the approaches to updating medical insurance catalogues vary considerably, each with distinct advantages and limitations. In the United States, decisions are heavily influenced by market dynamics and strategies of commercial insurance companies, which are highly adaptive to market changes but may at times prioritise profit over fairness, leading to inequities in healthcare provision \cite{Manning1984}. In contrast, the United Kingdom's National Health Service (NHS) utilises guidelines set by the National Institute for Health and Care Excellence (NICE) to base decisions on clinical and cost-effectiveness, striving to maximise public health benefits within constrained budgets. However, this approach can slow the adoption of new treatments due to its deliberate pace. Canada's decision-making process involves collaboration between multiple levels of government and professional review teams to balance costs and patient welfare, although the process can be lengthy and involve multiple levels of approval \cite{Parker1997}. Germany employs an independent joint assessment committee to ensure transparency and multi-stakeholder participation in its medical insurance catalogue decisions, although this process can be influenced by political and economic factors.

These various methods underscore a common set of challenges in balancing cost, efficiency, and fairness. The introduction of blockchain technology \cite{Vergne2020, liang2023, Amrendra2023, Cheng2018, Almasoud2020, Fernandes2023} could provide new technical means to address these issues, particularly in enhancing transparency and participation, reducing administrative costs, and increasing decision-making efficiency. Blockchain technology's advantages in data processing and confidentiality offer a viable solution by creating a transparent and automated voting and decision-making system that not only speeds up the process but also reduces human error and bias \cite{Kuo2017, Yaqoob2022, Catalini2020, Yakubov2018, Abdullah2017, Atzori2016, Wen2024, Huckle2016, Wang2019}. Additionally, smart contracts in the blockchain system can automatically execute decisions based on voting results, further enhancing the system's responsiveness and operational efficiency \cite{Kuo2017}. Introducing an economic incentive mechanism for doctors as voting nodes, especially providing rewards based on treatment outcomes and patient health results, encourages doctors to make decisions based on objective outcomes, promoting rational and effective allocation of medical insurance resources, thereby fundamentally enhancing the fairness and sustainability of the insurance system. This study proposes the conceptual model of such a medical insurance catalogue voting system.

\section{Methodology} \label{Basic Model}

In this section, we introduce briefly the model we are using as in \cite{Wen2024}. For the convenience of discussion, we call each doctor as an agent who is expected to vote for the choice of medicine to be paid by the national health insurance. In turn to agents' voting, they obtain credit points that represents the cumulative reputation or correctness, whose criteria will be defined later.


\subsection{Model of Voting System}
\label{subsec:On-chain}

Our voting operates in stages, concluding with assessments at each stage's completion. This staged approach is perfectly aligned with blockchain for potential implementation, where each "stage" mirrors a "block" in the blockchain.

\begin{figure}[h!]
\centering
\includegraphics[width=1.0\textwidth]{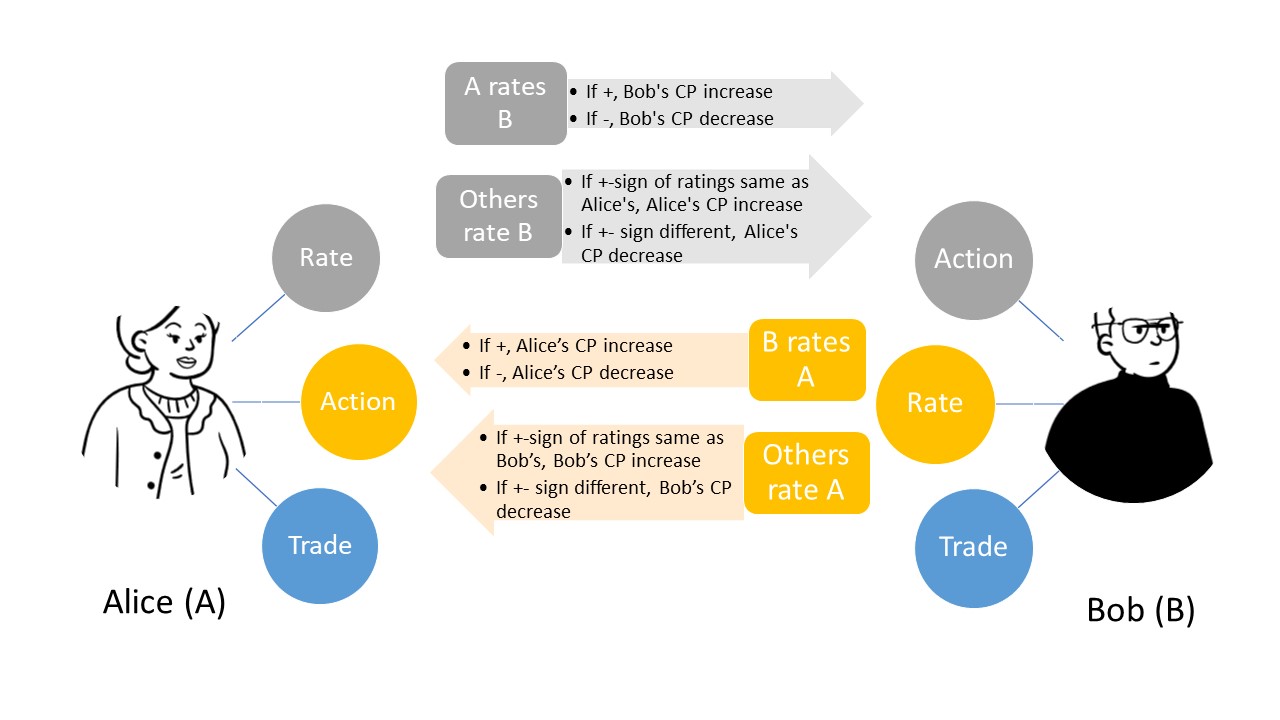}
\caption{Standard interaction stage in the doctor voting system: Doctor Alice and others evaluate Doctor Bob's Action, while Bob and others evaluate Alice's Action}
\label{fig:A typical round of doctor agents interactions}
\end{figure}

A reciprocal evaluation system is established within the voting market with potential to be implemented with blockchain technology, where each agent's reputation is logged as distinct credit points that evolve based on peer evaluations, as shown in Figure \ref{fig:A typical round of doctor agents interactions}. Agents like Alice and Bob engage in activities, with their actions affecting their credit scores through peer evaluations.

Credit scores that are publicly accessible guide trading decisions, creating significant motivation to secure positive evaluations. The prospect of potential future returns drives the need for precise assessments; congruent evaluations with subsequent comments enhance rewards, whereas discrepancies lead to penalties. The stakes involved alter the effect of evaluations, either amplifying or reducing previous assessments. This dynamic prompts agents to adjust their strategies in line with the prevailing consensus, mirroring typical market reactions to reputation and brand perception.

\subsection{Agents Representing Doctors Who Vote to Update Medical Insurance Catalogues} \label{Connecting the real world}

Participants are inclined to offer ratings anticipating future benefits. This is especially prevalent during the initial phase. Although speculators may buy credit points expecting them to rise in value, they have minimal financial incentive to skew ratings, thus preserving system integrity.

Moving to practical deployment, we outline a simple model categorising agents in the system:
\bigskip
\\
C1. Agents with Actions\\
C2. Rating Participants\\
C3. Investors
\bigskip

Agents might participate in one or more categories.

C1 agents, affected by potential gains or losses from ratings, tend to react quicker if they have staked considerable credit points. Visible staking draws consumers to high-staking producers, thereby encouraging new users to purchase credit points. Producers can also rate consumers, fostering overall responsible actions. C1 agents choose their staking amount—anything from zero to their entire wallet balance minus transaction fees.

C2 agents determine their staking based on how confident they feel about their ratings, motivated to rate honestly due to possible gains or losses aligned with other ratings.

C3 agents, engaging primarily in buying or selling credit points, don't impact system functionality but may boost credit point demand during initial stages through their purchases.

\subsection{Incentive Mechanisms of the Gains or Losses for Any Doctor Agent} \label{Assumptions of the Gains or Losses for Any Agent}

Let's consider agent \(i\) as a typical system participant, classified within three roles: C1 (action agents), C2 (rating agents), and C3 (investors). We note that trader agents (C3) don't impact the credit-related gains or losses of other agents, focusing our discussion on C1 and C2 categories.

Agent \(i\) can function as either a producer or consumer (C1), where \(S^A_{i_m}\) indicates the credit points staked for a specific action \(i_m\). Conversely, \(S^R_{ik_l}\) marks the credit points staked for evaluating the action \(k_l\) performed by another agent \(k\), affecting the staking outcomes positively or negatively.

The impact of any agent \(j\) rating the action \(i_m\) performed by agent \(i\) depends on the credit points both have staked and is given by:

\begin{eqnarray}
	\displaystyle {\Delta CP_{iji_m}^A} &=&\displaystyle {C_{R2A}S^A_{i_m}S^R_{ji_m},} 
	\label{momeqn1CPA} 
\end{eqnarray}

where \(\Delta CP_{iji_m}^A\) represents the variation in credit points for agent \(i\) due to action \(i_m\) as affected by agent \(j\), and \(C_{R2A}\) serves as the coefficient that connects ratings to actions.

Concerning the evaluations \(n_i\) made by agent \(i\) on action \(k_l\) executed by \(k\), the impact on profit or loss by any rating agent \(j\) remains proportional to the amount of credit points staked:

\begin{eqnarray}
	\displaystyle {\Delta CP_{ijk_l}^R} &=&\displaystyle {C_{R2R}S^R_{ik_l}S^R_{jk_l},} 
	\label{G_AiR} 
\end{eqnarray}

where \(C_{R2R}\) is the coefficient influencing the effect of one rating on another.

Summing up the inputs from all agents, the overall gains or losses for agent \(i\) are articulated as:

\begin{eqnarray}
	\displaystyle {\Delta CP_i} &=&
	\displaystyle {\Sigma_{j,m} C_{R2A}S^A_{i_m}S^R_{j_m} + \Sigma_{j,k,l} C_{R2R}S^R_{ik_l}S^R_{jk_l}},
	\label{momeqn1CPi} 
\end{eqnarray}

where \(\Delta CP_i\) denotes the aggregate changes in credit points for agent \(i\).

Prospective values for \(C_{R2A}\) and \(C_{R2R}\) are derived based on the aggregate of absolute values of pertinent staked amounts:

\begin{eqnarray}
	\displaystyle {C_{R2A}} &=&
	\displaystyle {\frac{1}{\Sigma_{j,m} |S^A_{i_m}S^R_{ji_m}|} ,} 
	\label{momeqn1CR2Atemp} 
\end{eqnarray}

\begin{eqnarray}
	\displaystyle {C_{R2R}} &=&
	\displaystyle {\frac{1}{\Sigma_{j,k,l} |S^R_{ik_l}S^R_{jk_l}|}.} 
	\label{momeqn1CR2A} 
\end{eqnarray}

This adjustment guarantees that losses in credit points will not surpass an agent's overall assets, fostering enduring engagements within the voting system.

Based on the above model settings, by interpreting each agent as a doctor in the health system as concerned, this study develops a new medical insurance catalogue voting system, that can reach agreement of the choice of medicine to be paid by national health insurance. This will be seen from the Monte Carlo simulations by using the same methods and settings as in \cite{Wen2024}.


\section{Monte Carlo Simulations} \label{sec:Monte_Carlo}

The Monte Carlo simulation method \cite{MetropolisUlam1949, Luo2022, Fishman2013, Raychaudhuri2008, Devroye1986, Law2000, Rubinstein2007} is a probabilistic technique used to understand the impact of risk and uncertainty in prediction and forecasting models. For our medical insurance catalog voting system, this simulation helps in analyzing various scenarios where the inputs, such as doctors' preferences and biases towards specific medical treatments, are varied randomly to generate a range of possible outcomes. By performing numerous iterations, we can estimate the likelihood of different medical treatments being selected for inclusion in the insurance catalog based on their perceived effectiveness and cost-efficiency.

\subsection{Application in Medical Insurance Catalogue Voting System}
Utilizing the methodology developed from \cite{Wen2024}, the Monte Carlo simulation was applied to the medical insurance catalogue voting system designed for this study. Each simulation cycle represented a voting session where doctors evaluated and voted on numerous treatment options based on predefined criteria such as clinical effectiveness, cost, and patient outcomes. The random variations in voting preferences were introduced to mimic real-world discrepancies and personal biases among the doctors.

In the design of our voting system, doctor agents are distinguished by their participation and assessments across varying interaction stages.Each stage coincides with a system's time step, where all ratings and credit points are refreshed. Ratings either increase or reduce credit balances, influenced by the credit points staked for activities and evaluations. These staking amounts are fixed in the Non-learning model, however, can be adjusted in the learning model in future rounds to reflect changes in strategies or levels of confidence.

The algorithm for the implementation of our voting system model is introduced briefly with the following pseudo code.

\begin{algorithm}[H]
\DontPrintSemicolon
\SetAlgoLined
\SetKwInOut{Input}{input}
\SetKwInOut{Output}{output}
\SetKwFor{For}{for}{do}{end for}
\SetKwIF{If}{ElseIf}{Else}{if}{then}{else if}{else}{end if}
\SetKw{Continue}{continue}
\SetKw{Return}{return}

\caption{Monte Carlo Simulation of the voting system}

\Input{Number of doctor agentss $N$, Number of rounds $T$, Initial credit points distribution, ...}
\Output{Results of the simulation}

\BlankLine
\emph{Initialize parameters (number of doctor agentss, number of rounds, initial credit points distribution, etc.)}\;
\For{each round $r$}{
\For{each doctor agents $i$}{
        \eIf{doctor agents $i$ is to skip round}{
            continue to next doctor agents\;
        }{   }
\For{each other doctor agents $j$}{
        \eIf{doctor agents $j$ is to skip round}{
            continue to next doctor agents\;
        }{   }
\For{each other doctor agents $k$}{
        \eIf{doctor agents $k$ is to skip round}{
            continue to next doctor agents\;
        }{
            
        }
                      Calculate credit point changes for doctor agents $i$ based on ratings from others\;
Calculate credit point changes for each doctor agents $j$ based on $j$’s ratings to others\;
}
}
}
        Sum the credit point changes for each doctor agents, based on its ratings from others and to others in the current round\;
        Update the credit point for each doctor agents, and then the total credit points in the system in the current round\;
        Save/Record data for the current round\;
}
Output and illustrate the cumulative results of the analysis across multiple rounds

\end{algorithm}

\begin{figure}[h!]
  \centering
  \includegraphics[width=0.6\textwidth]{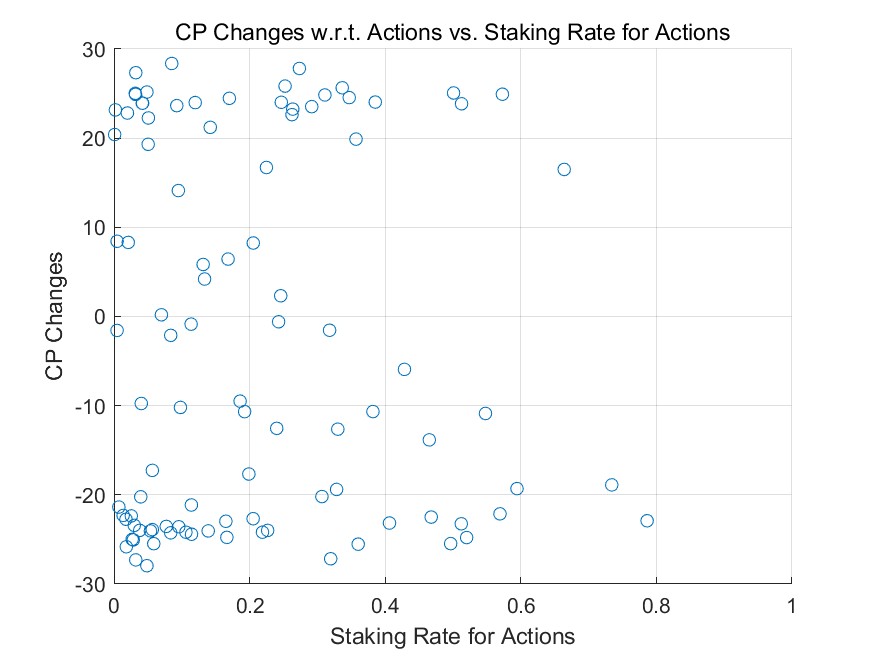} 
  \caption{Non-Learning Model: Changes of Credit Points over Staking Rate for Actions from doctor agents}
  \label{fig:Non-Learning Uniform with Consumer Selection}
\end{figure}

\begin{figure}[h!]
  \centering
  \includegraphics[width=0.6\textwidth]{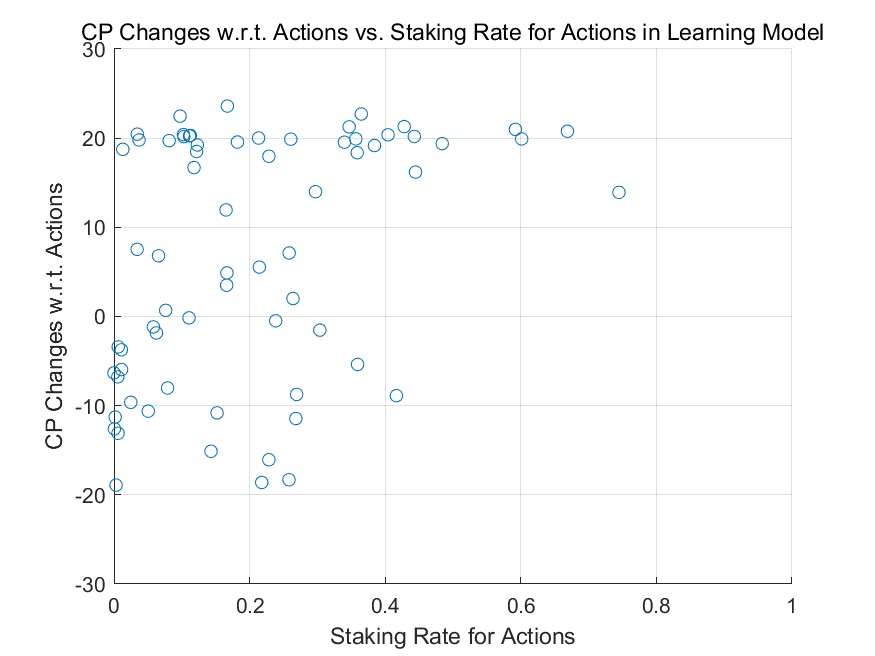} 
  \caption{Learning Model with Consumer Selection: Changes of Credit Points over Staking Rate for Actions from doctor agents}
  \label{fig:Learning Uniform with Consumer Selection}
\end{figure}

\subsection{Results}

As noted in \cite{Wen2024}, the transparent behaviour of doctor agents promotes consensus, which manifests as observable trends and likelihoods of specific medications or treatments being preferred, leading to updates in the medical insurance catalogue.
Figure \ref{fig:Non-Learning Uniform with Consumer Selection} shows that in the Non-Learning model, the incentive of the change of CP (credit points) is not sensitive to the staking rate for actions, which represents voting for specific medications or treatments. However,  Figure \ref{fig:Learning Uniform with Consumer Selection} shows that in the Learning model, the incentive of the change of CP (credit points) is positively correlated to the staking rate for actions, and hence, doctors are incentives to reach agreement in the voting for specific medications or treatments to be chosen or not for the updated medical insurance catalogue. Those who stake more will obtain more credit points in turn.

Economically, it's reasonable to suggest these doctor agents are more committed to their roles, possibly due to their expertise or other factors, often resulting in favourable evaluations. Conversely, less specialised doctor agents or those receiving consistently poorer reviews tend to become sidelined over time. This shift confirms the old saying: "the most specialised professionals undertake the most specialised duties."

\section{Conclusions} \label{sec:Conclusions}

Through the simulations, we are able to identify patterns and probabilities of certain medicines or treatments being favoured over others. These results were then compiled to propose an updated medical insurance catalogue. The simulations revealed a high degree of consensus on treatments that demonstrated clear clinical benefits and cost-effectiveness, whereas treatments with ambiguous or marginal benefits showed larger variations in acceptance.

The results from the Monte Carlo simulations provided us with a data-driven foundation to propose a medical insurance catalogue that not only reflects the collective expertise and judgement of the participating doctors but also aligns with the principles of cost-effectiveness and patient benefit. This approach significantly reduces the subjectivity and potential bias in deciding which treatments should be covered under medical insurance.

These insights are critical as they inform stakeholders about the potential impacts of different decision-making frameworks and support the implementation of a more robust, transparent, and equitable healthcare provisioning system.

\section{Further Investigation} \label{sec:Further Investigation}

\subsection{Blockchain Implementation and Voting Nodes}
To further investigate and validate the results from our Monte Carlo simulation, a blockchain-based system will be implemented where doctors act as nodes in the network. Each doctor node will participate in voting on the medical insurance catalog, ensuring transparency and traceability of decisions through blockchain's immutable ledger. This decentralized approach allows for a secure and verifiable record of all voting outcomes, which helps in assessing the alignment of the proposed catalog with actual preferences and practices within the medical community.

\subsection{Incentive Mechanism Based on Patient Survival Days}
An innovative incentive mechanism will be introduced, where doctors are rewarded based on the survival outcomes of patients who receive treatments they voted for. This aligns doctors' voting behaviors with patient health outcomes, encouraging choices that genuinely benefit patients. The effectiveness of this incentive system will be initially tested in a pilot program within selected regions or countries. The pilot will provide valuable insights into how such incentives impact decision-making and treatment outcomes.

\subsection{Pilot Testing and Optimization}
The pilot regions will serve as testbeds to fine-tune the voting system and the incentive mechanisms. By closely monitoring the pilot outcomes, we can make iterative improvements to the system, ensuring that it not only supports sustainable healthcare spending but also maximizes public health benefits. The goal is to refine the system to a point where it can be scaled up and potentially implemented on a national or international scale, providing a robust, optimized medical insurance catalog voting system that enhances both healthcare quality and economic efficiency.

The findings and methodologies from this further investigation will help in shaping a more effective and equitable healthcare system, where decisions are driven by data, secured by technology, and incentivized to promote the best possible patient outcomes.

\vspace{-6pt}



\bigskip
 
\bibliographystyle{unsrt}

\begin{thebibliography}{}

\bibitem[Walshe, 2009]{Walshe2009}
K. Walshe,
Pseudoinnovation: the development and spread of healthcare quality improvement methodologies,
\textit{International Journal for Quality in Health Care}, Volume 21, No. 3, 2009, pp. 153-159.

\bibitem[Kuo et al., 2017]{Kuo2017}
T.-T. Kuo, H.-E. Kim, and L. Ohno-Machado,
Blockchain distributed ledger technologies for biomedical and health care applications,
\textit{Journal of the American Medical Informatics Association}, Volume 24, No. 6, 2017, pp. 1211-1220.

\bibitem[Manning et al., 1984]{Manning1984}
Willard G. Manning, et al.,
A controlled trial of the effect of a prepaid group practice on use of services,
\textit{New England Journal of Medicine}, Volume 310, No. 23, 1984, pp. 1505-1510.


\bibitem[Parker and Wong, 1997]{Parker1997}
S.W. Parker and R. Wong,
Household income and health care expenditures in Mexico,
\textit{Health Policy}, Volume 40, No. 3, 1997, pp. 237-255.






\bibitem[Vergne, 2020]{Vergne2020}
JP Vergne,
Decentralized vs. Distributed Organization: Blockchain, Machine Learning and the Future of the Digital Platform,
\textit{Organization Theory}, Volume 1, Issue 4, October-December 2020, \url{https://doi.org/10.1177/2631787720977052}.


\bibitem[Liang, 2023]{liang2023}
Decui Liang, Yuanyuan Fu, Harish Garg,
A novel robustness PROMETHEE method by learning interactive criteria and historical information for blockchain technology-enhanced supplier selection,
\textit{Expert Systems with Applications, Volume 235, January 2024, 121107}.

\bibitem[Amrendra, 2023]{Amrendra2023}
Amrendra Singh Yadav, Vincent Charles, Dharen Kumar Pandey, Somya Gupta, Tatiana Gherman, Dharmender Singh Kushwaha,
Blockchain-based secure privacy-preserving vehicle accident and insurance registration,
\textit{Expert Systems with Applications, Volume 235, January 2024, 121107}.

\bibitem[Cheng, 2018]{Cheng2018}
Jiin-Chiou Cheng, Narn-Yih Lee, Chien Chi, Yi-Hua Chen,
Blockchain and smart contract for digital certificate,
\textit{2018 IEEE International Conference}.


\bibitem[Almasoud, 2020]{Almasoud2020}
Ahmed S. Almasoud, Farookh Khadeer Hussain, Omar K. Hussain,
Smart contracts for blockchain-based voting systems: A systematic literature review,
\textit{Journal of Network and Computer Applications}, Volume 170, 15 November 2020.

\bibitem[Fernandes, 2023]{Fernandes2023}
Claudio Piccolo Fernandes et al.,
A blockchain-based voting system for trusted VANET nodes,
\textit{Ad Hoc Networks}, Volume 140, 1 March 2023.

\bibitem[Yaqoob et al., 2022]{Yaqoob2022}
Yaqoob, I., Salah, K., Jayaraman, R., \& Al-Hammadi, Y. (2022). Blockchain for healthcare data management: opportunities, challenges, and future recommendations. \textit{Neural Computing and Applications}, 34, 11475–11490.

\bibitem[Catalini \& Gans, 2020]{Catalini2020}
Catalini, C., \& Gans, J.S. (2020). Some simple economics of the blockchain. \textit{Communications of the ACM}, 63(7), 80-90. Available at: \url{https://doi.org/10.1145/3359552}.

\bibitem[Yakubov et al., 2018]{Yakubov2018}
Yakubov, A., Shbair, W., Wallbom, A., et al. (2018). A Blockchain-Based PKI Management Framework. In \textit{The First IEEE/IFIP International Workshop on Managing and Managed by Blockchain} colocated with IEEE/IFIP NOMS 2018, Taipei, Taiwan, 23-27 April 2018. Available at: \url{https://hdl.handle.net/10993/35468}.

\bibitem[Abdullah et al., 2017]{Abdullah2017}
Abdullah, N., Hakansson, A., \& Moradian, E. (2017). Blockchain based approach to enhance big data authentication in distributed environment. In \textit{2017 Ninth International Conference}. IEEE. Available at: \url{https://ieeexplore.ieee.org/abstract/document/7993927}.

\bibitem[Atzori, 2016]{Atzori2016}
Atzori, M. (2016). Blockchain Technology and Decentralized Governance: Is the State Still Necessary? \textit{SSRN}. Available at: \url{https://www.ssrn.com/abstract=number_here}.





\bibitem[Wen et al., 2024]{Wen2024}
H. Wen, S. Sun, T. Huang, and D. Xiao,
An intrinsic integrity-driven rating model for a sustainable reputation system,
\textit{Expert Systems with Applications}, Volume 249, Part C, 2024, 123804.


\bibitem[Huckle et al., 2016]{Huckle2016}
S. Huckle, R. Bhattacharya, M. White, and N. Beloff,
Internet of things, blockchain and shared economy applications,
\textit{Procedia Computer Science}, Volume 98, 2016, pp. 461-466.

\bibitem[Wang et al., 2019]{Wang2019}
Y. Wang, J.H. Han, and P. Beynon-Davies,
Understanding blockchain technology for future supply chains: a systematic literature review and research agenda,
\textit{Supply Chain Management}, Volume 24, No. 1, 2019, pp. 62-84.
















\bibitem[Metropolis and Ulam, 1949]{MetropolisUlam1949}
Nicholas Metropolis and S. Ulam,
The Monte Carlo Method,
\textit{Journal of the American Statistical Association}, Volume 44, No. 247, Sep. 1949, pp. 335-341,
\url{http://www.jstor.org/stable/2280232}.


\bibitem[Luo et al., 2022]{Luo2022}
Changqi Luo, Behrooz Keshtegar, Shun Peng Zhu, Osman Taylan, Xiao-Peng Niu,
Hybrid enhanced Monte Carlo simulation coupled with advanced machine learning approach for accurate and efficient structural reliability analysis,
\textit{Comput. Methods Appl. Mech. Engrg.}, Volume 388, 2022, Article 114218,
\url{https://doi.org/10.1016/j.cma.2021.114218}.



\bibitem[Fishman, 2013]{Fishman2013}
George S. Fishman,
\textit{Discrete-Event Simulation: Modeling, Programming, and Analysis},
Springer Series in Operations Research and Financial Engineering, 2001.
\url{https://doi.org/10.1007/978-1-4757-3552-9}

\bibitem[Raychaudhuri, 2008]{Raychaudhuri2008}
Samik Raychaudhuri,
Introduction to Monte Carlo simulation,
In: \textit{2008 Winter Simulation Conference},
Date of Conference: 07-10 December 2008,
DOI: 10.1109/WSC.2008.4736059,
Publisher: IEEE.
\url{https://ieeexplore.ieee.org/abstract/document/4736059}

\bibitem[Devroye, 1986]{Devroye1986}
Luc Devroye,
\textit{Non-Uniform Random Variate Generation},
Springer-Verlag, 1986.
\url{https://doi.org/10.1007/978-1-4613-8643-8}

\bibitem[Law and Kelton, 2000]{Law2000}
Averill M. Law and W. David Kelton,
\textit{Simulation Modeling and Analysis},
McGraw-Hill, 3rd Edition, 2000.

\bibitem[Rubinstein and Kroese, 2007]{Rubinstein2007}
Reuven Y. Rubinstein and Dirk P. Kroese,
\textit{Simulation and the Monte Carlo Method},
Wiley, 2nd Edition, 2007.
\url{https://doi.org/10.1002/9780470181463}







\end{thebibliography}

\end{document}